\documentclass[12pt]{iopart}
\usepackage{appendix}
\usepackage{booktabs,tabularx,caption}
\usepackage[utf8]{inputenc}
\usepackage{graphicx}
\expandafter\let\csname equation*\endcsname\relax
\expandafter\let\csname endequation*\endcsname\relax
\usepackage{amsmath}
\usepackage{amssymb}
\usepackage{bm}
\usepackage{braket}
\usepackage{verbatim}

\begin{document}

\title{Rough 1D  Photonic Crystals: a transfer matrix approach}

\author{Leandro L Missoni$^1$,  Guillermo
  P Ortiz$^2$,  María Luz  Martínez Ricci$^1$, Victor J Toranzos$^3$ and W Luis Mochán$^4$}
\address{$^1$Instituto de Química, Física de los Materiales, Medioambiente y
  Energía (INQUIMAE-CONICET), DQIAQF, Facultad de Ciencias Exactas y
  Naturales, Universidad de Buenos Aires, Pabellón II, Ciudad Universitaria,
C1428EHA-Buenos Aires, Argentina.}
\address{$^2$Departamento de Física, Facultad de
  Ciencias Exactas Naturales y Agrimensura,  Universidad Nacional del
  Nordeste, Corrientes, Argentina}
\address{$^3$Departamento de Ingenieria, Facultad de
  Ciencias Exactas Naturales y Agrimensura,  Universidad Nacional del
  Nordeste, Corrientes, Argentina}
\address{$^4$Instituto de Ciencias F\'isicas,  Universidad Nacional
  Aut\'onoma de M\'exico, Av. Universidad s/n, Col. Chamilpa, 62210
  Cuernavaca, Morelos, M\'exico}
\ead{\mailto{missoni@qi.fcen.uba.ar}, \mailto{gortiz@exa.unne.edu.ar},
  \mailto{mricci@qi.fcen.uba.ar}, \mailto{mochan@fis.unam.mx}}

\begin{abstract}
  Interfacial roughness is prevalent in 1D photonic
  crystals and other layered structures, but is not generally accounted
  for in their design nor the analysis of their optical properties due to a lack of
  simple theoretical approaches. We present a transfer matrix formalism to incorporate
  the effects of interfacial roughness in the optical properties of
  stratified systems such as 1D photonic crystals and apply it to
  calculate the optical response of some nanoporous anodic
  alumina and porous silicon structures. We have validated our
  formalism by comparing our results to some experiments.
\end{abstract}
\noindent{\it Keywords\/}: roughness, transfer matrix, photonic crystals, optical
properties, porous alumina, porous silicon.

\submitto{\JOPT}

\section{Introduction}\label{sec:intro}

Nanomaterials have gained a lot of interest in diverse fields of
active research due to their unique
properties. Manipulation of their composition,
characteristic size, and shape allows tuning
their optical, electrical and chemical
properties, which might differ substantially from those of their
components. In the last decades, the variety of synthesis procedures for
nanomaterials has increased rapidly, and these structures can be found
in several configurations~\cite{Ding(2017)} such as
nanoparticles~\cite{Ahmadi(1996),Puntes(2001),Geonel(2002)},
aerogels~\cite{Anderson(1999)}, or thin films~\cite{Fuertes(2009)}.
In particular, thin film nanomaterials have gained a lot of interest
because they have the advantage of a simpler synthesis yielding a
robust platform for
design~\cite{Geist(2005),Martinez(2009),Fischereder(2012)} that allows
the fabrication of many nano-structures with diverse applications in
optoelectronics~\cite{Granqvist(2007)}, in solar
cells~\cite{Colodrero(2009)}, and even in portable nano-sensors using
effects such as surface enhanced Raman scattering (SERS)
~\cite{Wolosiuk(2014)}. Particularly interesting nano-systems are
one-dimensional photonic crystals (1D-PhC), periodic multilayered
systems that exhibit an optical band gap whose properties allow them
to be used as a sensing
platform~\cite{Nair(2010),Kumeria(2014)}, such as an
electromagnetic field assisted SERS sensor \cite{Luz(2017)}. Diverse
electromagnetic (EM) models have been implemented to understand
the optical response of the thin films, and
to obtain information from complex nanostructures embedded within the structure
taking advantage of non-invasive and non-destructive techniques such
as ultraviolet-visible
(UV-Vis) spectroscopy ~\cite{Sanchez(2013), Onna(2019)}. However,
while many synthesis methods yield fluctuations of the geometric
parameters of the nanostructures,
the roughness of the interfaces has usually not been taken into account
explicitly in the design or modeling of these
nano-systems.

Roughness has been widely studied in regards to several optical
phenomena\cite{Raether(1988)} and at different wavelength
scales. For one
or two slab problems with up to three interfaces  numerical methods
have been used to solve Maxwell
equations~\cite{Bourlier(2015)} and  asymptotic models for coherent
scattering have been applied~\cite{Pinel(2012)}. These have been used in geophysical problems
such as the study of reflections from geological layers of ground
penetration radar signals. Nevertheless, these
EM models are difficult to apply in multilayered problems. Roughness
has also been considered in numerical simulations using finite differences in the
time domain (FDTD) for 1D-PhC in the microwave region, and it has been
shown to produce appreciable red shifts of high frequency
features~\cite{Glushko(2008)}, a prediction that has been
experimentally confirmed. There have been
some efforts to model the coherent scattering for multilayered systems
with rough interfaces in the microwave region employing the Kirchhoff
approximation (KA)~\cite{Tabatabaeenejad(2013)}. Infinite 1D-PhC's composed of
non-dispersive, non-dissipative media with rough interfaces have also
been analyzed via 2D FDTD simulations~\cite{Maskaly(2004)}.  For the
case of wavelengths much larger than the period and for roughness
heights not exceeding some small fraction of the period, their
reflectance could be reproduced as well using an
homogenization procedure~\cite{Maskaly(2005)}.
Unfortunately these approaches do not take into
account energy losses through scattering. Diffuse
light scattering for surfaces with 1D roughness has also been
studied~\cite{Gonzalez-Alcalde(2016)}.
There is a lot of work done on roughness effects on wave propagation
\cite{Raether(1988),Ogilvy(1987),Voronovich(1998)}, but
many of
these EM models are not easily carried to the case of 1D-PhC.
Recently, a matrix formulation of the effects of
roughness~\cite{LujanCabrera(2019)} has been developed for
multilayered structures, but it cannot be easily
incorporated into the convenient and well known transfer matrix method
(TMM)~\cite{BornWolf(1999)} since it doesn't yield unimodular
matrices.
In this work we present a TMM for stratified systems, such as
1D-PhC, that incorporates the roughness of the interfaces under the
assumption of a small
angle condition~\cite{Elfouhaily(2004)}. Under this condition, the
scattering matrices \cite{Voronovich(1998)}
may be
appropriately averaged to yield {\em macroscopic} interfacial transfer matrices that
that can
be combined with those of other films and interfaces
to produce the full transfer matrix of the system, from which
all of its optical properties can be obtained readily. These transfer matrices are
consistent with the well
known~\cite{Ogilvy(1987),Voronovich(2007)} Kirchhoff approximation
(KA), whose validity has been widely discussed
~\cite{Pinel(2012),Tabatabaeenejad(2013),Sanchez-Gil(1995),Franco(2017)}.

To test our
formalism, we apply it to several systems and we compare its
predictions to some experimental results. We
first analyze the measured reflectance of a single
nanoporous anodic alumina (NAA) film. We have
selected NAA thin films due to their
reproducibility~\cite{Santos(2015)} and versatile control through
chemical
means~\cite{Chen(2015)} for diverse
applications~\cite{Santos(2014)}.
By applying periodic current or
voltage pulses during the anodizing process, NAA 1D-PhC may be
obtained~\cite{Nair(2010),Calvo(2016)}. Thus, we also study with our
formalism their
photonic band structure, their
reflectance and transmittance, and we compare the effects of constant roughness to
those of progressive roughness, dissipation, and thickness
fluctuations. As NAA 1D-PhC structures have been reported to yield the
best detection limits as sensor devices~\cite{Santos(2014),Chen(2015)}, for quantifying optically the presence of analytes~\cite{Chen(2015)Nanoscale}, we also analyze within our theory
experimental NAA 1D-PhC data, expecting that
incorporation of roughness considerations
may lead to better sensor designs.

The structure of the paper is the following. In Sec.~\ref{sec:teo} we
present our theory and we derive expressions for the transfer matrices
of rough interfaces whose profile is uncorrelated to those of their neighbors
interfaces, and we show how they can be incorporated into the usual
transfer matrix of a multilayered system. In \ref{app:correlated} we extend this theory to
account for the possibility of several mutually correlated interfaces.
In section~\ref{sec:1slab} we apply our formalism to study the
reflectance of a single rough slab and we make a detailed comparison
between the sample parameters obtained directly from SEM and AFM
images to those fitted from the optical reflectance using our model.
In  ~\ref{app:AI_sample} we provide details of the preparation
and characterization of our samples and of our optical measurements.
In section~\ref{sec:1DPhC} we study infinite and finite 1D-PhC's, obtain
their band structure and relate it to their reflectance. We
compare the effects of roughness to those of absorption and of thickness'
fluctuations. We use these results in Sec. \ref{sec:AII} where we
measure and model the reflectance spectrum of a finite NAA
1D-PhC. Finally, Sec.~\ref{sec:conclusion} is devoted to our
conclusion.

\section{Theory}\label{sec:teo}

In this section we develop first the usual theory of transfer
matrices for our stratified media with nominal flat interfaces in order to
establish our notation. Then, we will show how to
incorporate the roughness of the actual interfaces into the transfer
matrix formalism under some simplifying assumptions.

\subsection{Flat interfaces}\label{sec:flat-interfaces}

We
assume that the nominal system is time invariant and has translational
symmetry along the $xy$ plane, so that we can
consider fields with a well defined frequency $\omega$ and 2D wavevector $\bm Q$ along
$xy$.
For local isotropic media and for a
given TE or TM polarization, the transfer
matrix $\mathbf M(z_2, z_1)$ is a 2$\times$2 matrix that relates the components
of the electric field $E_{\parallel}$ and the magnetic field $H_{\parallel}$
parallel to the $xy$ plane,
and normal to the axis of the
structure which we take as the $z$ axis, evaluated at two arbitrary
heights $z_1$ and $z_2$,
\begin{equation}\label{eq:defM}
  \left(
  \begin{array}{c}
    E_\|\\H_\|
  \end{array}
  \right)_{z_2}
  =\mathbf M(z_2,z_1)
  \left(
  \begin{array}{c}
    E_\|\\H_\|
  \end{array}
  \right)_{z_1}.
\end{equation}
Several equivalent formulations have been proposed to obtain $\mathbf M$
\cite{Mochan(1987),Yeh(2005),Huerta(2018)}. Within a homogeneous
nonmagnetic layer $\alpha$ characterized by
a dielectric function
$\epsilon_\alpha$ and refractive index
$n_{\alpha}=\sqrt{\epsilon_{\alpha}}$, the field is in general the sum
of upward ($+$) and downward ($-$) going {\em plane} waves,
with wavevector components $k_\alpha^\pm = \pm k_\alpha$ along $z$ given by the dispersion relation
\begin{equation}\label{eq:kalpha}
  Q^2+k_\alpha^2=\epsilon_\alpha\frac{\omega^2}{c^2}.
\end{equation}
For economy in notation, we will use the nomenclature of plane waves
even if $Q$ is large, or $\epsilon$ is negative or complex, in which
case, $k_\alpha$ may acquire an imaginary part and the waves become
evanescent. We define $k_\alpha$ as the solution of
\eqref{eq:kalpha} with a positive imaginary
part. The surface impedance is defined as the quotient
\begin{equation}\label{eq:Z}
  Z= E_{\parallel}/H_{\parallel}.
\end{equation}
From Faraday's and Ampere-Maxwell's equations we obtain
\begin{equation}\label{ZTETM}
  Z^\pm_{\alpha}= \pm Z_\alpha=
  \pm \begin{cases}
    \frac{\omega}{k_{\alpha}c},&\text{(TE)}\\
    \frac{k_{\alpha}c}{\omega \epsilon_\alpha},&\text{(TM)}
  \end{cases}
\end{equation}
for upwards ($+$) and downwards ($-$) moving plane waves.
It is convenient to write $E_{\parallel}(z)$ and
$H_{\parallel}(z)$ in terms of the electric field amplitude of the
$\pm$ waves for TE polarization and in terms of the magnetic field
amplitudes for TM polarization. Denoting these amplitudes as
$\gamma^{\pm}(z)$ in both cases, we write
\begin{equation}\label{eq:EyHte}
    \left(
  \begin{array}{c}
    E_{\parallel}\\
    H_{\parallel}\\
  \end{array}
  \right)_z=
  \left(
  \begin{array}{cc}
    1&1\\
    Y_{\alpha}&-Y_{\alpha}\\
  \end{array}
  \right)
    \left(
  \begin{array}{c}
    \gamma^{+}\\
    \gamma^{-}\\
  \end{array}
  \right)_z,\quad\text{(TE)}
\end{equation}
and
\begin{equation}\label{eq:EyHtm}
    \left(
  \begin{array}{c}
    E_{\parallel}\\
    H_{\parallel}\\
  \end{array}
  \right)_z=
  \left(
  \begin{array}{cc}
    Z_{\alpha}&-Z_{\alpha}\\
    1&1\\
  \end{array}
  \right)
    \left(
  \begin{array}{c}
    \gamma^{+}\\
    \gamma^{-}\\
  \end{array}
  \right)_z.\quad\text{(TM)}
\end{equation}
where $Y_{\alpha}\equiv1/Z_{\alpha}$ is the surface admittance.
As $\gamma^\pm(z)$ propagate as plane waves for
both polarizations,
$\gamma^{\pm}(z)\propto\exp(\pm i
k_{\alpha}z)$ and
\begin{equation}\label{eq:BMpropagation}
     \left(
  \begin{array}{c}
    \gamma^{+}\\
    \gamma^{-}\\
  \end{array}
  \right)_{z_{b}}=
    \left(
  \begin{array}{cc}
    \exp(\rmi k_{\alpha}(z_{b}-z_{a}))&0\\
    0&\exp(-\rmi k_{\alpha}(z_{b}-z_{a}))\\
  \end{array}
  \right)
       \left(
  \begin{array}{c}
    \gamma^{+}\\
    \gamma^{-}\\
  \end{array}
  \right)_{z_{a}}
\end{equation}
for any $z_a$ and $z_b$ within the homogeneous layer.
Evaluating (\ref{eq:EyHte}) and (\ref{eq:EyHtm})  at
both nominal edges $z^{(0)}_{\alpha-1}$ and
$z^{(0)}_\alpha=z^{(0)}_{\alpha-1}+d_\alpha$ of the layer $\alpha$ of
width $d_\alpha$ (superscript $(0)$ denotes absence of roughness) and
using (\ref{eq:BMpropagation})
we obtain
\begin{equation}
      \left(
  \begin{array}{c}
    E_{\parallel}\\
    H_{\parallel}\\
  \end{array}
  \right)_{z^{(0)}_\alpha}=
  \mathbf M_\alpha^{(0)}      \left(
  \begin{array}{c}
    E_{\parallel}\\
    H_{\parallel}\\
  \end{array}
  \right)_{z^{(0)}_{\alpha-1}}
\end{equation}
for both TE and TM polarizations, where
\begin{equation}\label{eq:m}
\mathbf M_\alpha^{(0)}= \mathbf M_{\alpha}(d_\alpha)=\left(\begin{array}{cc}
    \cos(k_{\alpha}d_\alpha)& iZ_{\alpha}\sin(k_{\alpha}d_\alpha)\\
    iY_\alpha\sin(k_{\alpha}d_\alpha) &\cos(k_{\alpha}d_\alpha)\\
  \end{array}\right),
\end{equation}
is the transfer matrix of
a homogeneous layer $\alpha$ that transfers the fields across its
nominal width $d_\alpha$. We remark that the determinant of this matrix is $1$, as required to comply with time reversal symmetry.

The continuity of $E_\|$ and $H_\|$ imply that the
fields may be transferred from the ambient $\alpha=0$ towards the substrate
$\alpha=N+1$ through the layers $\alpha=1\ldots N$ by the
matrix
\begin{equation}\label{eq:nominal}
  \mathbf M^{(0)}=\mathbf M^{0}_{N}\mathbf M^{0}_{N-1}\ldots \mathbf M^{0}_2\mathbf M^{0}_1.\quad\text{(nominal)}
\end{equation}

\subsection{Rough interfaces}
\label{sec:roughness}
We consider a system as above, but in which a single interface
$\alpha$ separating medium $\alpha$ from $\alpha+1$ is displaced
from its nominal position at $z^{(0)}_\alpha$ to a new position
$z_\alpha=z^{(0)}_\alpha+\zeta_\alpha$. The transfer matrix of the modified system can then be
obtained by replacing $\mathbf M_\alpha^{(0)}$ and $\mathbf M_{\alpha+1}^{(0)}$ in
(\ref{eq:nominal}) by
$\mathbf M_\alpha(d_\alpha+\zeta_\alpha)=\mathbf M_\alpha(\zeta_\alpha)\mathbf M_\alpha^{(0)}$ and
$\mathbf M_{\alpha+1}(d_{\alpha+1}-\zeta_\alpha)=\mathbf M_{\alpha+1}^{(0)}\mathbf M_{\alpha+1}(-\zeta_\alpha)$
respectively. This is equivalent to the replacement of the product
$\mathbf M^{(0)}_{\alpha+1}\mathbf M^{(0)}_\alpha$ by $\mathbf M^{(0)}_{\alpha+1}\mathbf M^I_\alpha
\mathbf M^{(0)}_\alpha$, where we introduced an effective interface transfer
matrix
\begin{equation}\label{eq:interface}
  \mathbf M^I_\alpha=\mathbf M_{\alpha+1}(-\zeta_\alpha)\mathbf M_\alpha(\zeta_\alpha).
\end{equation}
Assuming now that all interfaces $\alpha=0\ldots N$ are shifted by a
corresponding displacement $\zeta_\alpha$, we
can build the complete transfer matrix as the product
\begin{equation}\label{eq:desplazada}
    \mathbf M=\mathbf M^I_N\mathbf M^{(0)}_{N}\mathbf M^I_{N-1}\mathbf
    M^{(0)}_{N-1}\mathbf M^I_{N-2}\ldots \mathbf M^I_2\mathbf
    M^{(0)}_2\mathbf M^I_1\mathbf M^{(0)}_1\mathbf
    M^I_0.\quad\text{(displaced)}
\end{equation}

\begin{figure}
  \begin{center}
    \includegraphics{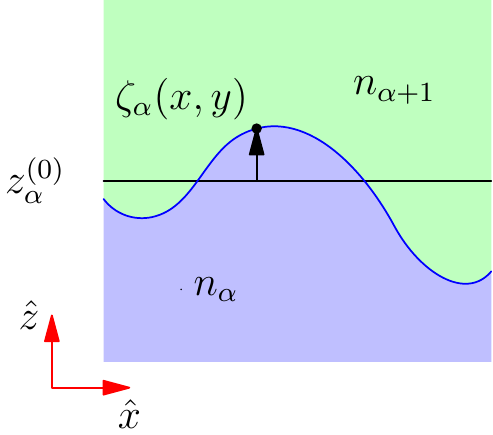}
    \caption{
      \label{fig:perfil}
      Rough surface separating two consecutive layers, $\alpha$ and
      $\alpha+1$, described by refractive indices $n_\alpha$ and
      $n_{\alpha+1}$. The interface is defined through a
      height function $\zeta_\alpha(x,y)=z_\alpha(x,y)-z^{(0)}_\alpha$ given by its
      position with respect to the nominal surface $z=z_\alpha$.
    }
  \end{center}
\end{figure}

Now we introduce the roughness through $x$ and $y$ dependent height profiles
$\zeta_\alpha(x,y)$, $\alpha=0\ldots N$, as illustrated in
figure \ref{fig:perfil}.
In the spirit of the well known Kirchhoff
approximation \cite{Ogilvy(1987),Voronovich(2007)} in the limit of low
angle roughness \cite{Eckart(1953)} it seems tempting to apply
(\ref{eq:desplazada}) for each  $x,y$ and average it over the $xy$
plane, or, for a randomly rough system, over all
realizations of an ensemble. For the case
of mutually uncorrelated profiles, this would be equivalent to a replacement of
$\mathbf M^I_\alpha$ by its average $\langle \mathbf M^I_\alpha\rangle$. Nevertheless, this would
be {\em wrong}; the transfer matrix {\em is not} a quantity that may
be meaningfully averaged,
as the determinant of the average doesn't agree in general with the
average of the determinant, and the transfer matrix {\em ought to be
unimodular} in order to be consistent with time inversion symmetry;
even  though
$\det \mathbf M^I_\alpha=\det \mathbf M_{\alpha+1}(-\zeta_\alpha)\det
\mathbf M_{\alpha}(\zeta_\alpha)=1$,
$\det \langle \mathbf M^I_\alpha\rangle =\det\langle
\mathbf M_{\alpha+1}(-\zeta_\alpha) \mathbf M_{\alpha}(\zeta_\alpha)\rangle\ne1$.
Nevertheless, the scattering matrix of each interface {\em is} a quantity
that may be safely and meaningfully averaged.

We define the scattering $\mathbf S_\alpha$
matrix for the $\alpha$-th interface as
\begin{equation}\label{eq:S}
  \mathbf S_\alpha=\left(
  \begin{array}{cc}
    r_{\alpha}^+&t_{\alpha}^-\\
    t_{\alpha}^+&r_{\alpha}^-
  \end{array}
  \right),
\end{equation}
as it produces the outgoing waves
$\gamma^-_\alpha(z)=O^-_\alpha \rme^{-\rmi k_\alpha (z-z_\alpha^{(0)})}$ and
$\gamma^+_{\alpha+1}(z)=O^+_\alpha \rme^{\rmi k_{\alpha+1}  (z-z_\alpha^{(0)})}$
when applied to the incoming waves $\gamma^+_\alpha(z)=I^+_\alpha
\rme^{\rmi k_\alpha (z-z_\alpha^{(0)})}$ and $\gamma^-_{\alpha+1}(z)=I^-_\alpha
\rme^{-\rmi k_{\alpha+1} (z-z_\alpha^{(0)})}$, namely,
\begin{equation}\label{eq:O=SI}
  \left(
  \begin{array}{c}
    O^-_\alpha\\O^+_\alpha
  \end{array}
  \right)
  =
  \mathbf S_\alpha
  \left(
  \begin{array}{c}
    I^+_\alpha\\I^-_\alpha
  \end{array}
  \right),
\end{equation}
where we defined the amplitudes $O^\pm_\alpha$ and $I^\pm_\alpha$
using the nominal plane $z=z_\alpha^{(0)}$ as a reference, and where
$r^\pm_\alpha$ and $t^\pm_\alpha$ are the reflection and transmission
amplitudes corresponding to waves that impinge on the interface
$\alpha$ moving upwards ($+$) within layer $\alpha$ or moving
downwards ($-$) within layer $\alpha+1$.  We may characterize the
fields above and below the interface in terms of the
components of $\mathbf S_\alpha$ through the
matrices
\begin{equation}\label{eq:RTE}
  \mathbf A_\alpha=
  \left(
  \begin{array}{cc}
    t^+_\alpha&1+r^-_\alpha\\
    Y_{\alpha+1}t^+_\alpha&-Y_{\alpha+1}(1-r^-_\alpha)
  \end{array}
\right)
\quad\text{(TE)}
\end{equation}
and
\begin{equation}\label{eq:LTE}
  \mathbf B_\alpha=
  \left(
  \begin{array}{cc}
    1+r^+_\alpha&t^-_\alpha\\
    Y_\alpha(1-r^+_\alpha)&-Y_\alpha t^-_\alpha
  \end{array}
  \right)\quad\text{(TE)}
\end{equation}
for TE polarization, and
\begin{equation}\label{eq:RTM}
  \mathbf A_\alpha=
  \left(
  \begin{array}{cc}
    Z_{\alpha+1} t^+_\alpha&-Z_{\alpha+1}(1-r^-_\alpha)\\
    t^+_\alpha&1+r^-_\alpha
  \end{array}
  \right)\quad\text{(TM)}
\end{equation}
and
\begin{equation}\label{eq:LTM}
  \mathbf B_\alpha=
  \left(
  \begin{array}{cc}
    Z_\alpha(1-r^+_\alpha)&-Z_\alpha t^-_\alpha\\
    1+r^+_\alpha&t^-_\alpha
  \end{array}
  \right)\quad\text{(TM)}
\end{equation}
for TM polarization. These matrices yield the electromagnetic field
extrapolated to each side, above and below the nominal surface when they act on the
incoming amplitudes $(I_\alpha^+, I_\alpha^-)^T$.
Thus, we may relate $\mathbf A_\alpha$ and $\mathbf B_\alpha$ through
$\mathbf M^I_\alpha$ and write
\begin{equation}\label{eq:MvsS}
  \mathbf M_\alpha^I=\mathbf A_\alpha \mathbf B_\alpha^{-1}.
\end{equation}

For a flat displaced interface we can obtain the
elements of the scattering matrix either from
Eqs. (\ref{eq:interface}) and (\ref{eq:MvsS}) or through a simple
extrapolation of the fields from the actual towards the nominal interface,
\begin{equation}\label{eq:Svszeta}
  \mathbf S_\alpha
  =
  \left(
    \begin{array}{cc}
      r^{(0)+}_\alpha \rme^{2\rmi k_\alpha\zeta_\alpha}&t^{(0)-}_\alpha
                                                  \rme^{-\rmi\Delta k_\alpha\zeta_\alpha}\\
      t^{(0)+}_\alpha \rme^{-\rmi\Delta k_\alpha\zeta_\alpha}&r^{(0)-}_\alpha
                                                         \rme^{-2\rmi k_{\alpha+1}\zeta_\alpha}
    \end{array}
  \right)
\end{equation}
where $r^{(0)\pm}_\alpha$ and $t^{(0)\pm}_\alpha$ are the Fresnel
coefficients of the nominal interface \cite{Hecht(1986)}
and $\Delta k_\alpha=k_{\alpha+1}-k_\alpha$. Thus, not unexpectedly,
the effect of a rigid shift of an interface is to incorporate
simple {\em phase} factors proportional to $\zeta_\alpha$ into the
optical coefficients.  The interfacial
matrix \eqref{eq:interface} can be recovered in this case from
Eqs. \eqref{eq:RTE}-\eqref{eq:Svszeta}.

Going back to a rough interfacce, in the spirit
of the Kirchhoff approximation for low angle surfaces
\cite{Ogilvy(1987),Voronovich(1998)}, we average
(\ref{eq:Svszeta}),
\begin{equation}\label{eq:Sav}
  \langle \mathbf S_\alpha \rangle \equiv
  \left(
    \begin{array}{cc}
      \Braket{r^+_\alpha}&\Braket{t^-_\alpha}\\
      \Braket{t^+_\alpha}&\Braket{r^-_\alpha}
    \end{array}
  \right)
  =
  \left(
  \begin{array}{cc}
    r^{(0)+}_\alpha \Braket{\rme^{2\rmi k_\alpha\zeta_\alpha}} &t^{(0)-}_\alpha
    \Braket{\rme^{-\rmi \Delta k_\alpha\zeta_\alpha}} \\
    t^{(0)+}_\alpha \Braket{\rme^{-\rmi\Delta k_\alpha\zeta_\alpha}} &r^{(0)-}_\alpha
      \Braket{\rme^{-2\rmi k_{\alpha+1}\zeta_\alpha}}
  \end{array}
  \right),
\end{equation}
we replace the optical coefficients by their averages in
Eqs. (\ref{eq:RTE})-(\ref{eq:LTM}) to obtain $\langle \mathbf B_\alpha\rangle$ and
$\langle \mathbf A_\alpha\rangle$ and transferring the averaged fields across the
nominal interface we obtain, in analogy with (\ref{eq:MvsS}),
the {\em macroscopic} interface transfer matrix $\mathbf M^{MI}_\alpha$ as
\begin{equation}\label{eq:MMI}
  \mathbf M_\alpha^{MI}=\langle \mathbf A_\alpha\rangle\langle \mathbf
  B_\alpha\rangle^{-1}.
\end{equation}

From Eqs. (\ref{eq:RTE}), (\ref{eq:LTE}) and (\ref{eq:Sav}), we notice
that for TE polarization
\begin{equation}\label{eq:detRTE}
  \det \langle \mathbf A_\alpha\rangle= -2 Y_{\alpha+1} t_\alpha^{(0)+}
  \langle \rme^{-\rmi\Delta k_\alpha\zeta_\alpha}\rangle
\end{equation}
and
\begin{equation}\label{eq:detLTE}
  \det \langle \mathbf B_\alpha\rangle= -2 Y_{\alpha} t_\alpha^{(0)-}
  \langle \rme^{-\rmi\Delta k_\alpha\zeta_\alpha}\rangle.
\end{equation}
Nevertheless, from Fresnel's formulae we obtain
\begin{equation}\label{eq:Zt=Zt}
  Y_\alpha t_\alpha^{(0)-}=Y_{\alpha+1} t_\alpha^{(0)+},
\end{equation}
so that $\det \mathbf A_\alpha=\det \mathbf B_\alpha$ and from
(\ref{eq:MMI}) we verify the required unimodularity condition
\begin{equation}\label{eq:detMMI=1}
  \det \mathbf M^{MI}_\alpha=1.
\end{equation}
It is trivially verified that this condition also holds for TM polarization.

Finally, we get a {\em macroscopic} unimodular transfer matrix
$\mathbf M^M$for
the complete
system by replacing all the interface transfer matrices in
(\ref{eq:desplazada}) by their macroscopic counterparts
(\ref{eq:MMI}),
\begin{equation}\label{eq:MM}
    \mathbf M^M = \mathbf M^{MI}_N \mathbf M^{(0)}_{N} \mathbf
    M^{MI}_{N-1} \mathbf M^{(0)}_{N-1} \mathbf M^{MI}_{N-2}\ldots
    \mathbf M^{MI}_2 \mathbf M^{(0)}_2\mathbf M^{MI}_1\mathbf
    M^{(0)}_1\mathbf M^{MI}_0.
\end{equation}
We remark that although our formulation is close to that of
\cite{LujanCabrera(2019)}, their $\mathbf W$ matrix is not
unimodular.

From the transfer matrix of the system we may obtain its optical
properties straightforwardly using standard procedures. For example, the
reflection and transmission amplitudes, $r$ and $t$, when the system is illuminated
from the ambient towards the substrate, may be obtained by solving the
$2\times2$ system of equations
\begin{equation}\label{eq:rytTEslab}
\left(\begin{array}{c}
   t\\
   Y_{N+1} t\\
\end{array}\right) =\mathbf M^M
\left(\begin{array}{c}
  1+r\\
  Y_0(1-r)\\
  \end{array}\right), \quad(\mathrm{TE})
\end{equation}
and
\begin{equation}\label{eq:rytTMslab}
\left(\begin{array}{c}
   Z_{N+1}t\\
   t\\
\end{array}\right) =\mathbf M^M
\left(\begin{array}{c}
  Z_0(1-r)\\
  1+r\\
  \end{array}\right). \quad(\mathrm{TM})
\end{equation}

Naturally, we define the nominal positions $z^{(0)}_\alpha$ by demanding $\langle
\zeta_\alpha\rangle=0$. Thus, in the limit of small height roughness  we
may approximate (\ref{eq:Sav}) to order $\zeta_\alpha^2$ by
\begin{equation}\label{eq:Savsmall}
  \langle \mathbf S_\alpha \rangle =
  \left(
  \begin{array}{cc}
    r^{(0)+}_\alpha (1-2k_\alpha^2 \tilde \zeta_\alpha^2) &t^{(0)-}_\alpha
    (1-\frac{1}{2} (\Delta k_\alpha)^2 \tilde \zeta_\alpha^2 ) \\
    t^{(0)+}_\alpha (1-\frac{1}{2} (\Delta k_\alpha)^2
    \tilde \zeta_\alpha^2) &r^{(0)-}_\alpha
    (1-2k_{\alpha+1}^2 \tilde \zeta_\alpha^2)
  \end{array}
  \right),
\end{equation}
where $\tilde\zeta_\alpha\equiv\langle\zeta_\alpha^2\rangle^{1/2}$ is the RMS height,
so that all optical coefficients are reduced by a factor of order
$\tilde \zeta_\alpha^2/\lambda^2$, with $\lambda$ the free
space wavelength. This wavelength-dependent reduction is due to the
energy that is lost through scattering out of the specular direction
and plays a role analogous but not identical to
dissipation \cite{Dorado(2009)}.
On the other hand, if $\zeta_\alpha$ is not necessarily small but
obeys a Gaussian distribution with zero mean, we may evaluate
(\ref{eq:Sav}) to obtain
\begin{equation}\label{eq:SavG}
  \langle \mathbf S_\alpha \rangle =
  \left(
  \begin{array}{cc}
    r^{(0)+}_\alpha \rme^{-2k_\alpha^2\tilde\zeta_\alpha^2} &t^{(0)-}_\alpha
    \rme^{-(\Delta k_\alpha)^2\tilde\zeta_\alpha^2/2} \\
    t^{(0)+}_\alpha \rme^{-(\Delta k_\alpha)^2\tilde\zeta_\alpha^2/2} &r^{(0)-}_\alpha
    \rme^{-2k_{\alpha+1}^2\tilde\zeta_\alpha^2}
  \end{array}
  \right).
\end{equation}
A simple substitution of any of these into (\ref{eq:MMI}) yields
the corresponding interface matrix $\mathbf M^{MI}_\alpha$. Thus, in
the small roughness case \eqref{eq:Savsmall} we
obtain
\begin{equation}\label{eq:Mr}
  \mathbf M_\alpha^{MI}=\left(\begin{array}{cc}
   1 + \frac{(Z_{\alpha+1} - Z_\alpha)(k_\alpha + k_{\alpha+1})^2}
   {2(Z_\alpha + Z_{\alpha+1})}\tilde\zeta_\alpha^2
   & -\frac{(Z_\alpha k_\alpha - Z_{\alpha+1}
     k_{\alpha+1})^2} {(Z_\alpha +
     Z_{\alpha+1})}\tilde\zeta_\alpha^2\\ -\frac{(Z_{\alpha+1}
     k_\alpha - Z_\alpha k_{\alpha+1})^2} {(Z_\alpha + Z_{\alpha+1}) Z_\alpha
     Z_{\alpha+1}}\tilde\zeta_\alpha^2 &
   1 - \frac{(Z_{\alpha+1} - Z_\alpha) (k_\alpha + k_{\alpha+1})^2}
   {2(Z_\alpha + Z_{\alpha+1})}\tilde\zeta_\alpha^2\\
 \end{array}\right),
 \end{equation}
and in the general case, we obtain
\begin{align}\label{eq:MrG}
  \mathbf M_\alpha^{MI}=&
  \frac{\braket{t_\alpha^+}}{2}\left(
  \begin{array}{cc}
    1 & Z_{\alpha}\\
    Y_{\alpha+1} &Z_{\alpha}Y_{\alpha+1}\\
  \end{array}\right)\\
  &+
  \frac{1}{2\braket{t_\alpha^-}}\left(\begin{array}{cc}
    (1+\braket{r_\alpha^-})(1-\braket{r_\alpha^+}) & -(1+\braket{r_\alpha^-})(1+\braket{r_\alpha^+}) Z_{\alpha}\\
     -(1-\braket{r_\alpha^-})(1-\braket{r_\alpha^+})Y_{\alpha+1} & (1-\braket{r_\alpha^-})(1+\braket{r_\alpha^+}) Z_{\alpha}Y_{\alpha+1}\\
 \end{array}\right).\nonumber
\end{align}
For a Gaussian roughness we can further substitute
(\ref{eq:SavG}).

In summary, we have developed a formalism that allows us to calculate
the macroscopic effective transfer matrix of a multilayered system with
rough interfaces using the Kirchhoff small-angle approximation. The
transfer matrix is the product of the usual transfer matrices for each
layer, corresponding to their nominal width, alternating with
interface transfer matrices which can be obtained from the average of
the corresponding scattering matrices. We obtained explicit
expressions in terms of the variance of the height of the interface
for the case of small height roughness and for the case of Gaussian
roughness.  In the following sections, we will illustrate
the use of these matrices to calculate the optical properties of some
rough stratified systems and we will compare some of our results to
experiment. In \ref{app:correlated} we extend the above theory to
account for the possibility of several mutually correlated interfaces.

\section{Application I: Nanoporous anodic alumina single rough
  film} \label{sec:1slab}

In this section, we first test the proposed model on a single rough
film of nanoporous anodic alumina (NAA) over an Al substrate. The sample
preparation and the experimental procedures are described in detail in
~\ref{app:AI_sample}. We measured with an atomic force microscope
(AFM)
the surface profile $\zeta$, its RMS mean $\tilde\zeta$, the profile
slope $s=|\nabla_\parallel \zeta|$
  and its RMS mean $\tilde s$,
averaging over several positions on
each sample. All
studied samples exhibited $\tilde s< 0.02$, so that the small angle
approximation holds, giving us confidence in the use of the Kirchhoff
approximation.

Table~\ref{tabla1} summarizes the preparation and properties of 3 samples:
an electropolished Al surface (S1), as those used as substrates for S2
and S3 which correspond to NAA films grown under different anodizing
conditions.
\begin{table}
  \setlength{\tabcolsep}{3pt}
  \small
  \begin{tabular}{c|cccc|ccc|ccc}
 \toprule
Sample& \multicolumn{4}{c|}{Anodization} &
 \multicolumn{3}{c|}{Characterization} & \multicolumn{3}{c}{Fit}\\
    \midrule
      & $t$ [s] & $V$ & $J$ & Electrolyte & $\tilde\zeta^{*}$ [nm] &
     $d^{\dagger}$ [nm] & $P^{\dagger}$ [\%] & $\tilde\zeta$ [nm] & $d$ [nm] & $P$ [\%] \\
    \midrule
    S1 & 300 & 10&  & HClO$_4$ & 2 & - & - & 1.5 & - & - \\
    S2 & 30 && 11.2 & H$_2$SO$_4$ & 5 & 260 & 15 & 5.8 & 278 & 20\\
    S3 & 30 && 44.8 & H$_2$SO$_4$ & 6 & 1020 & 30 & 5.1 & 1130 & 30\\
    \bottomrule
 \end{tabular}
 \caption{Analysis of 3 samples: S1 is an electropolished Al
   substrate. S2 and S3 are NAA films on Al prepared through different
   anodizing conditions. $t$ denotes the anodizing time,
   $V$ or $J$ denote either the voltage (in volts) or the
   current density (in mA/cm$^2$) applied in the corresponding
   {\it electrolyte} (see \ref{app:AI_sample} for details).
   $\tilde\zeta$ is the RMS roughness, $d$ the thickness of the film, and
   $P$ the alumina porosity. The characterization was done through
   AFM (indicated by $*$) or SEM ($\dagger$). The fit was done by modeling the
   reflectance.}
 \label{tabla1}
\end{table}
The samples were characterized using a scanning electron microscope (SEM)
to measure their thickness $d$ and porosity $P$ and an atomic force
microscope (AFM) to determine their
roughness. The uncertainties are around 10-15\% for parameters defined by the SEM and AFM
measurements.

\begin{figure}
  \begin{center}
      \includegraphics[width=0.9\textwidth]{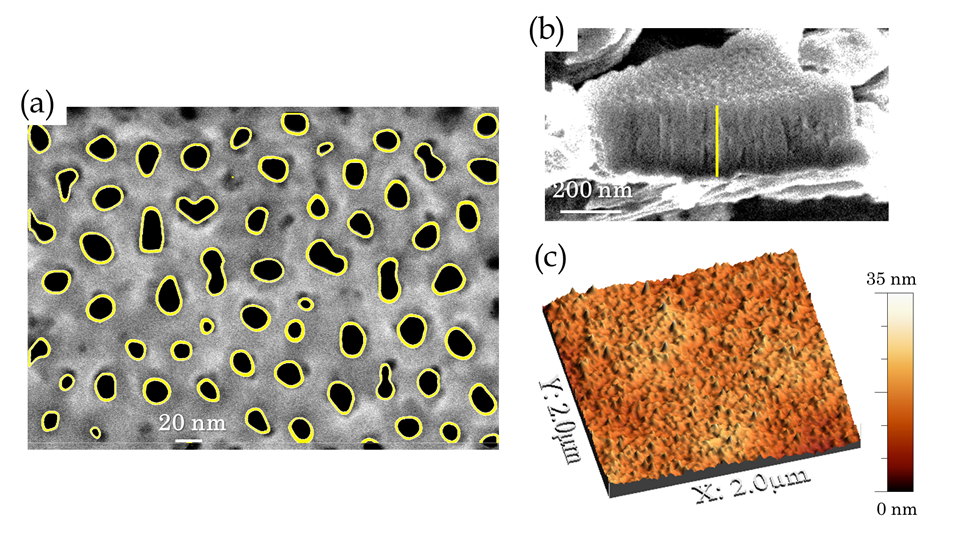}
  \caption{\label{fig:SEM+AFM} Some of the images used to characterize
    the sample S2. (a) Top-view SEM image with a porosity map
    superimposed (scale bar is 20 nm). (b) Side-view SEM showing the
    thickness of the film (scale bar is 200nm). (c) AFM image of the
    surface (the vertical color code from 0 to 35 nm is indicated).}
  \end{center}
\end{figure}

As an example, in figure \ref{fig:SEM+AFM} we show some of
the analyzed images for the sample S2, including top- and side-view SEM
micro-graphs and an AFM height map.
We developed a computational code in the PERL/PDL
language~\cite{pdl2010} to identify and measure the pores from our SEM
images and draw their porosity maps. In figure \ref{fig:SEM+AFM}a we
superposed the porosity map on the SEM top view image. We obtained the
thickness from the side view (figure \ref{fig:SEM+AFM}b) using the ImageJ
package \cite{schneider2012nih}. Notice that our sample has texture on
two scales, as can be observed in the AFM image
(figure \ref{fig:SEM+AFM}c). The smallest one corresponds to the pores,
and we deal
with it by homogenizing it using an effective medium theory
\cite{Ortiz(2018)}. The remaining texture is dealt through our transfer
matrix model for rough layers.

\begin{figure}
  \includegraphics[width=0.9\textwidth]{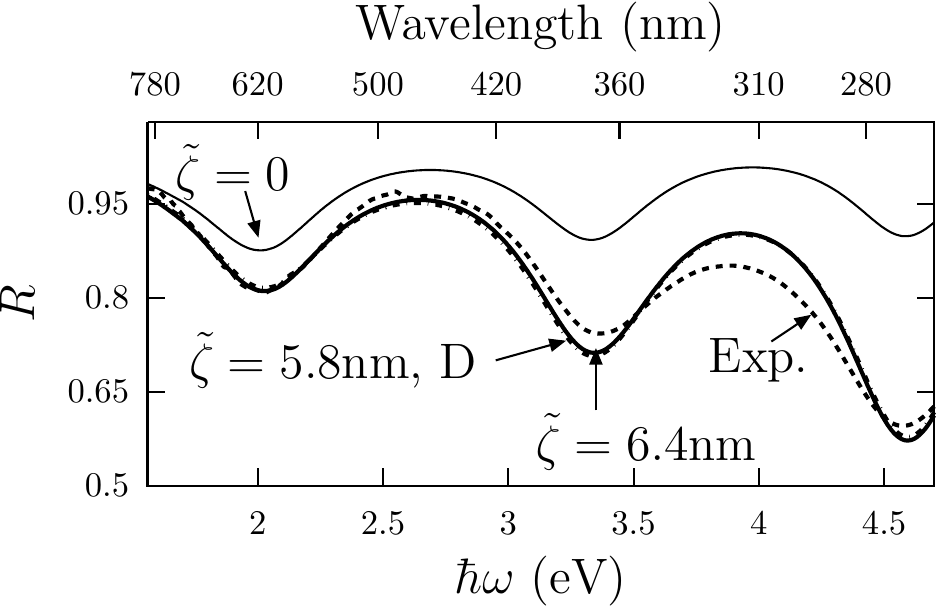}
  \caption{\label{fig:singleSlab}
    Normal incidence reflectance $R$ of sample S2 normalized to the
    reflectance $R_0$ of the substrate as a
    function of frequency and wavelength. We show experimental results
    (dashed), and several theoretical results fitted to experiment
    through the width $d$, porosity $P$, and/or roughness height
    $\tilde\zeta_0=\tilde\zeta_1=\tilde\zeta$. We show results for a
    flat film without dissipation ($d=274$nm, $P=18\%$,
    $\tilde\zeta=0$, thin solid), a rough film without dissipation ($d=278$, $P=20\%$,
    $\tilde\zeta=6.4$nm, thick solid), and
    a rough film with dissipation (D) due to the presence of
    $\text{Fe}_2\text{O}_3$ and CuO impurities (see text, $d=278$nm, $P=20\%$,
    $\tilde\zeta=5.8$nm, dash-dotted).}
\end{figure}

We measured the reflectance as
described in ~\ref{app:AI_sample} for the 3 samples described
in Table~\ref{tabla1}. We fitted the results using (\ref{eq:MM}) and the Gaussian roughness model.
We obtained the
reflection amplitude $r$ and the reflectance $R=|r|^2$ from
Eqs.~\eqref{eq:rytTEslab} and \eqref{eq:rytTMslab}. As an example of
our fitting procedure,
in figure \ref{fig:singleSlab} we show the normal-incidence reflectance
$R$ measurement for sample S2 as a function
of frequency and wavelength and the corresponding fits under
different assumptions. To perform any of the calculations
we need the refractive index for all the relevant media. The substrate is
Al \cite{Palik(1985)} while the film is NAA. The
refractive index of the alumina phase was taken from
reference~\cite{polyanskiy(2020)}, but to account for the nanoporous
character of the film we took the porosity $P$ as an adjustable
parameter and we applied Bruggeman's
2D~\cite{Ortiz(2018)} effective media theory to obtain the index of
the film. This is a very simple model but is
satisfactory enough for the system under study. The fitting parameters
were the RMS heights $\tilde\zeta_0=\tilde\zeta_1=\tilde\zeta$ of both
interfaces (which we assumed to be equal), the thickness $d$ and the porosity $P$. We
obtained these three parameters using the MINUIT
optimization package~\cite{Minuit(1994)}.
Figure \ref{fig:singleSlab} shows that in the absence of roughness the calculated
reflectance displays interference oscillations with a separation that
is simply related to
the nominal optical thickness of the film. Our experimental results display
these oscillations, but they also show a decrease in the value of the reflectance as the
frequency increases. This decay is also displayed by the fitted curve
that incorporates roughness, yielding a fitted roughness of
$\tilde\zeta=6.5$nm. If we also account for dissipation within the
alumina, due to the presence of impurities of $\text{Fe}_2\text{O}_3$ and CuO
with concentrations $0.13\%$ and $0.02\%$ respectively, as corresponds
to Al 1100~\cite{Aluminio1100} we obtain slightly smaller fitted roughness
$\tilde\zeta=5.8$nm.

These results may
be qualitatively understood through the wavelength dependent
scattering out of the specular direction that is incorporated
implicitly within our model. As discussed above
(\ref{eq:Savsmall}), this effect is of order
$(\tilde\zeta/\lambda)^2$.

The agreement between theory and experiment for the other samples
is similar to that shown in figure \ref{fig:singleSlab} for sample
S2. All the fitted parameters are
also included in Table~\ref{tabla1} and for the NAA slabs correspond to the dissipative case.  Notice the good agreement between the fitted parameters and the corresponding parameters obtained from AFM and SEM. For the case of the fitted
parameters, the uncertainty is around $2\%$ taking into account experimental data from different sample scanned regions.

\section{One Dimensional Photonic Crystal}\label{sec:1DPhC}

We consider a 1D photonic crystal (1D PhC) consisting of
the periodic alternation of two layers of refractive indices $n_1$
and $n_2$, nominal thicknesses $d_1$ and $d_2$ and period $L=d_1+d_2$.  We assume their
interfaces are rough, described by small mutually uncorrelated height functions
with corresponding RMS values $\tilde\zeta_1$
and $\tilde\zeta_2$.
Our
formulation above allows us to incorporate the effects of
roughness into an effective macroscopic transfer matrix for one period
by simple matrix
multiplication of the appropriate interface matrices, such as
\begin{equation}\label{eq:1DPhCM}
  \mathbf M_L^M=
  \mathbf M^{MI}_2
  \mathbf M^{(0)}_2
  \mathbf M^{MI}_1
  \mathbf M^{(0)}_1.
\end{equation}
The normal modes of the system are Bloch waves which are easily
obtained from the transfer matrix by solving the eigenvalue problem
\begin{equation}\label{eq:eigen}
  \mathbf M_L^M
      \left(
  \begin{array}{c}
    E_{\parallel}\\
    H_{\parallel}\\
  \end{array}
  \right)_{nL}=
  \Lambda
  \left(
  \begin{array}{c}
    E_{\parallel}\\
    H_{\parallel}\\
  \end{array}
  \right)_{nL},
\end{equation}
to obtain the eigenvalues
\begin{equation}\label{eq:eigenval}
  \Lambda_{\pm}=\exp(\pm \rmi KL),
\end{equation}
and eigenvectors
\begin{equation}\label{eq:eigenvec}
  \left(
  \begin{array}{c}
    E_{\parallel}\\
    H_{\parallel}\\
  \end{array}
  \right)_{nL}\propto
  \left(
  \begin{array}{c}
    Z_{\pm}\\
    1\\
  \end{array}
  \right),
\end{equation}
where $K$ is the Bloch vector along the axis of the structure
and $Z_\pm$ are the surface impedances for Bloch waves propagating (or
decaying) towards $\pm z$ evaluated at the position $nL$,
$n=1,2\ldots$ corresponding to the even numbered interfaces. As the
transfer matrix is unimodular, we can
rewrite the dispersion relation as
\begin{equation}\label{eq:KL}
  \cos(KL)=(m_{11}+m_{22})/2,
\end{equation}
where $m_{ij}$, $i,j=1,2$ are the matrix elements of $\mathbf M_L^M$.
The impedances are
\begin{equation}\label{eq:impedance}
  Z_{\pm}=\frac{m_{12}}{\exp(\pm \rmi KL)-m_{11}}=\frac{\exp(\pm \rmi KL)-m_{22}}{m_{21}}
\end{equation}

\subsection{Photonic Bands}\label{sec:resbbandas}
\begin{figure}
  \begin{center}
  \includegraphics[width=\textwidth]{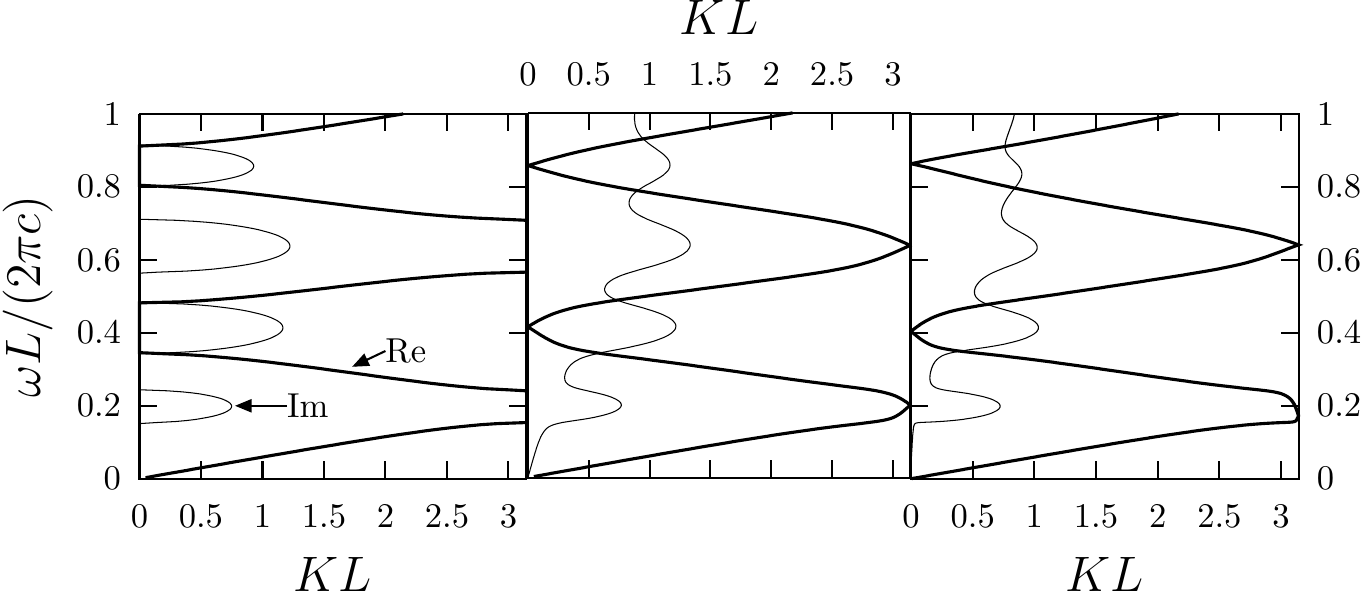}
  \caption{\label{fig:fig1} Dispersion relation of the Bloch modes
    propagating along the axis of an infinite 1D-PhC made of alternating layers of
    nominal widths $d_1=100$nm and $d_2=80$nm.
    We show the real (thick) and
    imaginary (thin)  parts of the
    normalized Bloch wave vector $KL$, with $L$ the period, as a function of the normalized
    frequency $\omega L/2\pi c$ for a dispersionless and dissipationless
    system with flat interfaces with real dielectric constants
    $\epsilon_1=12$ and $\epsilon_2=1$ (left), for the same system but
    with some dissipation, with $\epsilon_1=12+1.2i$, $\epsilon_2=1+0.1i$
    (middle)
    and for the system without dissipation but with Gaussian rough
    surfaces, with $\epsilon_1=12$, $\epsilon_2=1$,
    $\tilde\zeta_1=10$nm and $\tilde\zeta_2=8$nm (right).}
  \end{center}
\end{figure}

In figure \ref{fig:fig1} we show the dispersion relation $\omega$ vs. $K$
of the Bloch waves of an infinite 1D-PhC, calculated as discussed
above for parallel wavevector $Q=0$.
The system consists of two alternating films with thicknesses $d_{1}$=100nm and
$d_{2}$=80nm. To elucidate the role of roughness and of dissipation,
we calculated the modes for three cases: dispersionless, dissipationless media with
flat surfaces ($\epsilon_1=12$, $\epsilon_2=1$, $\tilde
\zeta_1=\tilde\zeta_2=0$), the same system with some dissipation
artificially added ($\epsilon_1=12+1.2i$, $\epsilon_2=1+0.1i$, $\tilde
\zeta_1=\tilde\zeta_2=0$), and the same system without dissipation but
with some Gaussian roughness  ($\epsilon_1=12$, $\epsilon_2=1$, $\tilde\zeta_1=8$nm,
$\tilde\zeta_2=10$nm).
For the dissipative case we added an imaginary
part to $\epsilon_\alpha$ equal to 10\% of its real part. For the rough
case, we chose the roughness amplitude as 10\% of the width of the
corresponding layer.
Notice that this roughness is not small for the
highest frequencies in the figure. Thus, \eqref{eq:Mr} is not
applicable, but we use \eqref{eq:MrG}.
For the flat system without dissipation (figure \ref{fig:fig1}, left) we
found, as expected, bands for which
the modes may propagate, with $\text{Im}\,K=0$ and with alternating positive
and negative group velocity, and gaps for which propagation is
forbidden and the {\em modes} decay exponentially, with $\text{Im}\,K\ne
0$ and $\text{Re}\,K=0$ or $\pm \pi/L$, at the center or at the edges
of the Brillouin zone. The center of the gaps
are at frequencies $\omega_m$ whose corresponding free-space
wavelength $\lambda_m=2\pi c/\omega_m$ is a sub-multiple of twice the
optical width of a period, yielding constructive interference of the
waves back-scattered by successive periods,
\begin{equation}\label{eq:Constructive}
n_{1}d_{1} +n_{2}d_{2}=m \lambda_m/2,
\end{equation}
and where $\text{Im}\,K$ peaks.
In the case with dissipation (figure \ref{fig:fig1}, center), $\text{Im}\,K$ also displays maxima corresponding
to the gaps, but $\text{Re} K$ for those frequencies is no longer a constant
at the center or at the edges of the Brillouin zone,
corresponding to both
decay and propagation. There are also regions where $\text{Im}K$ becomes
relatively small, corresponding to the propagating bands, but it is not
zero, so there is some decay due to extinction through absorption. For
high enough frequencies the minima
of $\text{Im}K$ is almost $1/L$, so that the decay distance becomes of the
order of one period. Of course, these are not actual modes that may be
excited within an
infinite system, but they may be excited in truncated photonic
crystals or in crystals with defects.
In the case with roughness (figure \ref{fig:fig1}, right), the results are very similar to those
obtained in the case with dissipation, as roughness also yields
extinction, though its origin is scattering out of the specular
direction instead of absorption.  We remark that we didn't calculate
explicitly the non-specular scattered fields, but the optical theorem
implies that the energy they carry away manifests itself through
destructive interference in the specular term
\cite{BornWolf(1999)}.

\begin{figure}
  \begin{center}
  \includegraphics[width=0.8\textwidth]{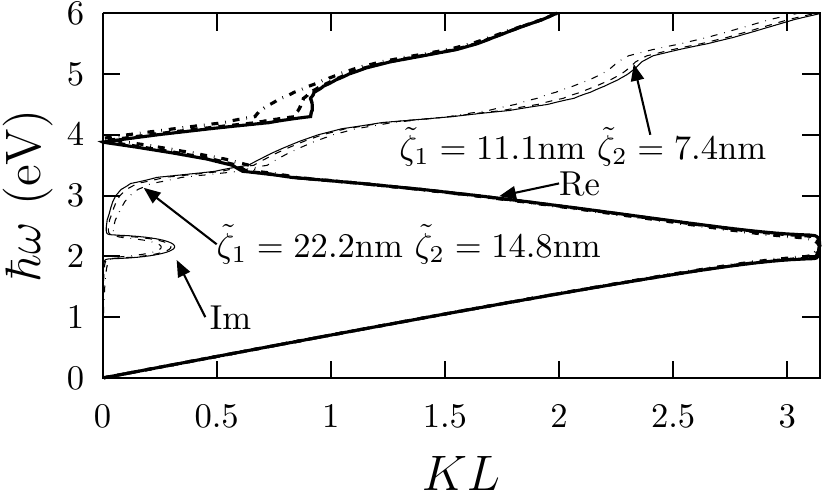}
  \caption{\label{fig:fig1SP} Band diagrams of Bloch modes propagating
    along the axis of a 1D-PhC made of nanoporous anodic silicon (NAS)
    layers of width $d_1=111$ nm with 73\%
    porosity alternating with NAS layers of width $d_2= 74$ nm with
    54\% porosity. We show the real part (thick) and imaginary part
    (thin) of the normalized Bloch's vector $KL$ as a function of
    photon energy $\hbar\omega$. We show results for flat
    interfaces (solid), for rough interfaces with RMS
    heights $\tilde\zeta_{1}=11.1$nm and
    $\tilde\zeta_{2}=7.4$nm (dashed), and for rough interfaces with
    $\tilde\zeta_{1}=22.2$nm and
    $\tilde\zeta_{2}=14.8$nm (dot dashed).}
  \end{center}
\end{figure}
In order to explore roughness effects for realistic
systems, in
figure \ref{fig:fig1SP} we show the band structure for a nanoporous
anodic silicon (NAS) 1D-PhC made of two alternating layers with
porosities $p_1=73$\% and $p_2=54$\% and widths $d_1=111$nm and
$d_2=74$ nm, as obtained from \cite{Toranzos(2008)a}.
We obtained $\epsilon_\alpha$ from the porosity
$p_\alpha$ and the dielectric function of silicon, taken from Palik's
handbook \cite{Palik(1985)}, using the Bruggeman 2D model
\cite{Ortiz(2018)}.  In figure \ref{fig:fig1SP} we show results
corresponding to a flat structure and to a structure with rough
surfaces considering two cases, a RMS height of 10\% of the corresponding layer's
width, i.e., $\tilde\zeta_1=11.1$ nm and
$\tilde\zeta_2=7.4$nm, and a RMS height of 20\% of the layer's
width ($\tilde\zeta_1=22.2$ nm and
$\tilde\zeta_2=14.8$nm). As expected, the middle
of the first band gap at $\hbar\omega_1=2.24$eV complies with
(\ref{eq:Constructive}), with $m=1$.  For this system, the
effects of dissipation dominate those of roughness. For the 10\%
case the roughness corrections are barely visible, though for the 20\%
they are relatively small but clearly discernible. We
expect roughness to be more important for materials with a smaller
dissipation, such as NAA.

\begin{figure}
  \begin{center}
  \includegraphics[width=0.8\textwidth]{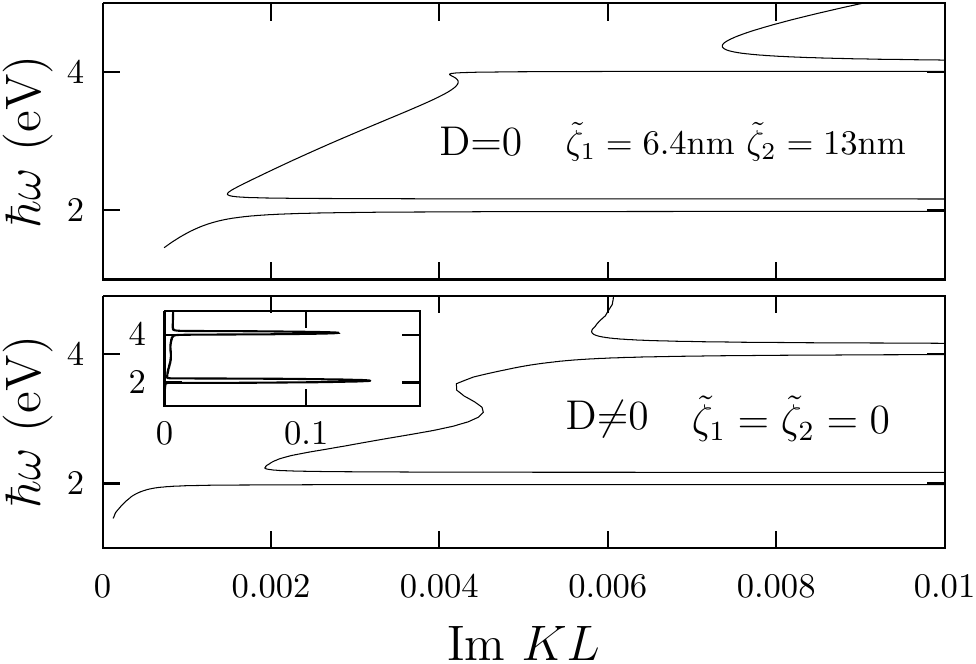}
  \caption{\label{fig:fig1AP} Band structure for a 1D-PhC of two
    alternating layers of NAA with porosities $p_1=7\%$ and
    $p_2=35\%$, and widths $d_1=64$ nm and $d_2= 130$ nm. We show the
    the normalized imaginary part of the Bloch vector $\text{Im}\,KL$
    as a function of energy $\hbar\omega$.  We
    present results corresponding to a dissipationless system with
    rough interfaces ($D=0$, $\tilde\zeta_1=6.4$ nm,
    $\tilde\zeta_2=13$nm, top), and for a system with dissipation
    corresponding to $\text{Fe}_2\text{O}_3$ and CuO impurities, as in
    Sec. \ref{sec:1slab}, but with flat interfaces (D$\ne$0,
    $\tilde\zeta_1=\tilde\zeta_2=0$, bottom).
    The inset corresponds to a change of scale.
  }
  \end{center}
\end{figure}
In figure \ref{fig:fig1AP}
we show the band diagram of an NAA 1D-PhC made up two alternating layers
of widths $d_1=64$ nm and $d_2=130$ nm and porosities $p_1=7\%$
and $p_2=35\%$. We used
Bruggeman's 2D model in order to calculate the dielectric functions
$\epsilon_\alpha$ ($\alpha=1, 2$) of the porous layers from that of Al
\cite{Aluminio1100,Palik(1985)}.
The calculation was done both for the case
of flat interfaces but including dissipation within the alumina due to impurities of
$\text{Fe}_2\text{O}_3$ and CuO, as in Sec. \ref{sec:1slab}, and for
that of  rough interfaces
characterized by $\tilde\zeta_1=6.4$ nm and $\tilde\zeta_2=13$nm (so that
$\tilde\zeta_\alpha/d_\alpha=0.1$) and without dissipation within
the alumina. We plot the imaginary part of the Bloch vector $K$.
Two band gaps are
clearly seen in the band structure, consistent with
(\ref{eq:Constructive}), in which the imaginary part of
Bloch's vector $K$ has large peaks (see inset of figure \ref{fig:fig1AP})
with or without roughness of the order of $0.1/L$, corresponding to
small penetration depths of $\approx10L$. Besides the gaps, there are
propagation bands for which $\text{Im}\,K$ would be null in the
absence of roughness or dissipation. Nevertheless,
$\text{Im}
K\ne0$ for the rough system and takes values as high as
$K\approx0.005/L$ corresponding to a penetration distance
of at most $\approx 200$ periods. For the parameters chosen, the
effects of roughness are qualitatively similar to those of dissipation.

\subsection{Finite 1D-PhC}\label{sec:resFinite}

Now, we consider a finite stratified system of width $NL$ made from
$N$ repetitions of a unit cell composed of two layers of nominal thicknesses $d_1$ and
$d_2=L-d_1$, refractive indices $n_1$ and $n_2$, and with mutually uncorrelated
rough surfaces characterized by the RMS heights $\tilde \zeta_1$ and
$\tilde \zeta_2$, respectively. The refractive index of ambient is
$n_0$ while for the substrate is $n_{2N+1}=n_s$.
We can obtain the reflectance and transmittance of the system from
Eqs. (\ref{eq:MM})-(\ref{eq:rytTMslab}).

\begin{figure}
  \begin{center}
    \includegraphics[width=0.46\textwidth]{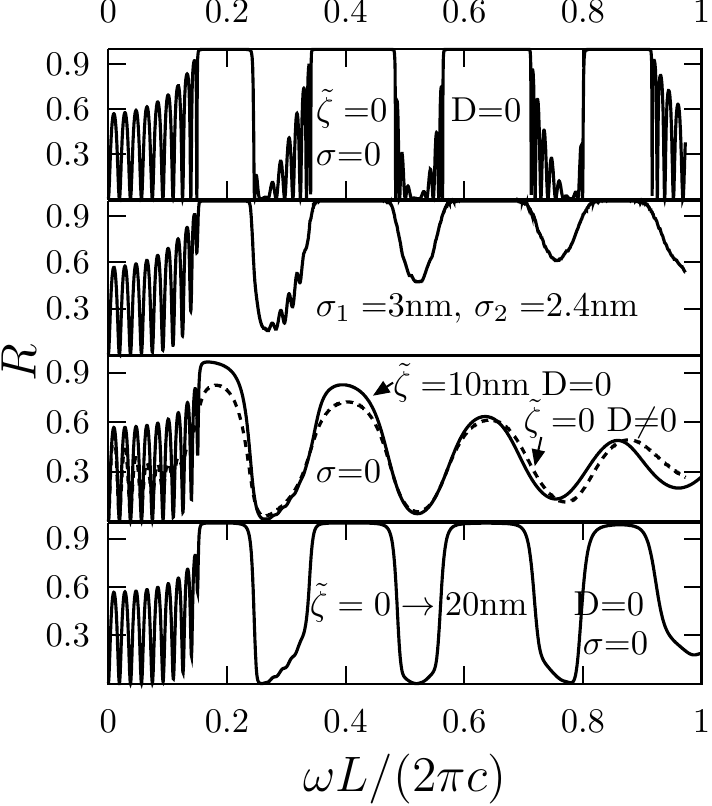}%
    \includegraphics[width=0.46\textwidth]{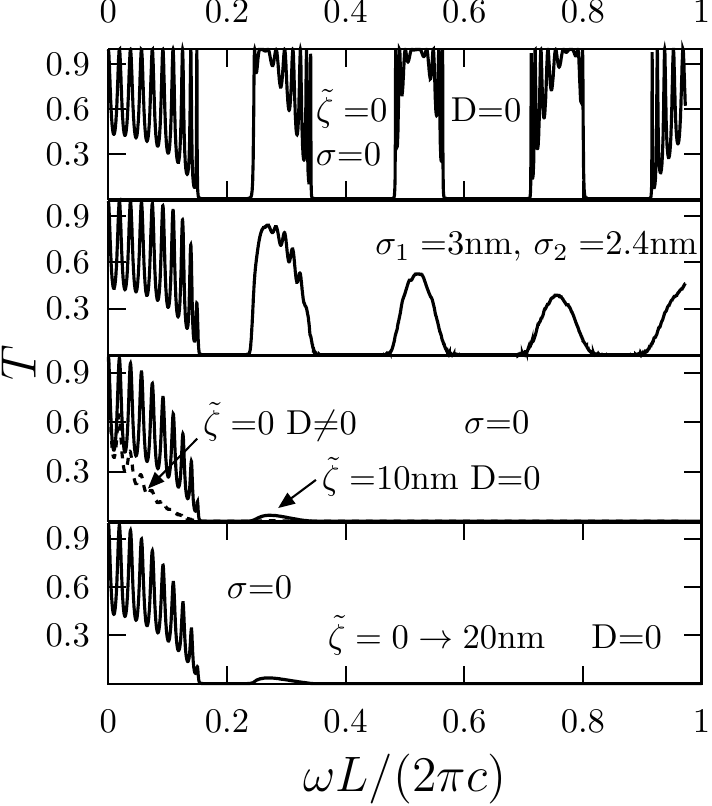}
  \caption{\label{fig:progreRugo}  Normal incidence reflectance $R$
    (left) and transmittance $T$ (right) of a free-standing
    ($n_0=n_s=1$) finite 1D-PhC made up of $N=10$ pairs of films of widths
    $d_1=100$nm and $d_2=80$nm as a function of the normalized
    frequency $\omega L/2\pi c$ where $L=d_1+d_2$. We consider several
    cases: no dissipation ($D=0$) corresponding to the real
    dielectric functions $\epsilon_1=12$ and $\epsilon_2=1$, with no
    roughness $\tilde\zeta_1=\tilde\zeta_2=0$ and no fluctuations
    $\sigma_1=\sigma_2=0$ (top row); no
    dissipation, no roughness and width fluctuations
    $\sigma_1=3$nm, $\sigma_2=2.4$nm (second row); roughness without
    dissipation nor fluctuations, $\tilde\zeta_1=\tilde\zeta_2=10$nm,
    $D=0$ and $\sigma_1=\sigma_2=0$ (solid) and dissipation with no
    roughness nor fluctuations, $\epsilon_1=12+1.2i$,
    $\epsilon_2=1+0.1i$, $\tilde\zeta_1=\tilde\zeta_2=0$ and
    $\sigma_1=\sigma_2=0$ (dashed) (third row); roughness growing
    linearly with height $\zeta_1=\zeta_2=0\to20$nm without dissipation
    nor fluctuations (bottom row).
  }
  \end{center}
\end{figure}

In figure \ref{fig:progreRugo} we show the normal incidence reflectance
$R$ and transmittance $T$ spectra of a free standing finite 1D-PhC in air
($n_0=n_s=1$) consisting of $N=10$ periods of the same photonic
crystals as in figure \ref{fig:fig1}, with and without dissipation and
roughness. We also present results for a system
with flat surfaces and no roughness nor dissipation, but whose thicknesses
have mutually uncorrelated fluctuations, and
for a system without dissipation nor width fluctuations but with a roughness whose
amplitudes grow linearly from the surface
towards the substrate. As shown in the top row of figure \ref{fig:progreRugo}, in
the case without
dissipation, roughness nor fluctuations, there are relatively wide
regions of very high reflectance $R\approx1$ and very small
transmittance $T\approx 0$
corresponding to the band gaps shown in figure \ref{fig:fig1}. Outside
of these regions, the reflectance and transmittance show strong
oscillations due to the interference of the waves reflected from the
front and back surfaces of the system. These oscillations take $R$ to
values as low as 0. Correspondingly, $T$ takes values as high as 1.

In the second row of figure \ref{fig:progreRugo}, we show the reflectance
and transmittance averaged over an ensemble of 1D-PhC's as those corresponding to the first row, but
with fluctuations in the widths of each film obeying a Gaussian
distribution with standard deviations $\sigma_1=3$nm and
$\sigma_2=2.4$nm corresponding to $3\%$ of their nominal
widths. We used one thousand ensemble members for our calculation.
The purpose of this calculation is to allow for a finite transverse
coherence length $\xi$ \cite{BornWolf(1999)} for the incident light, so that
the contributions to the reflectance and transmittance from
illuminated regions farther apart than $\xi$ add incoherently.
This situation will prove relevant for the system studied in the next section.

The results show regions where $R\approx1$ and $T\approx0$ as in the
top row, but the oscillating structure due to the interference is
damped by the averaging process, as the frequencies corresponding to
each maxima and minima differ for each member of the
ensemble. This damping is larger at higher frequencies, for which the
oscillations displayed by the first row become narrower and their
frequency spacing becomes smaller.
The third row of figure \ref{fig:progreRugo} shows the effects of
roughness characterized by the amplitudes $\tilde\zeta_1=\tilde\zeta_2=10$nm. In
this case, $R$ shows maxima corresponding to the band
gaps of figure \ref{fig:fig1}, but they don't attain the value 1,
as in the top and second rows. The reason is that in this case,
energy is lost through scattering, reducing the specular reflectance.
In our treatment of roughness, we averaged
scattering matrices, thus
incorporating phase information that is not present in our treatment
of thickness fluctuations corresponding to the second row of
figure \ref{fig:progreRugo}, for which we averaged the reflectance and
transmittance. Hence the qualitative differences.
Corresponding to the maxima in $R$ for the case of a rough surface
there are minima in $T$. Nevertheless, most of the energy is reflected
or scattered away so that beyond $\omega\approx0.3\times 2\pi c/L$ the
transmittance becomes negligible. The third row of figure \ref{fig:progreRugo}
also shows results for
a flat absorptive system with $\epsilon_1=12+1.2i$ and
$\epsilon_2=1+0.1i$. The result of dissipation is similar to that of roughness
without dissipation.

In the bottom row of figure \ref{fig:progreRugo} we show the effect of a
roughness that increases linearly in height from the front
($\tilde\zeta=0$) towards the back ($\tilde\zeta=20$nm) so that its
average coincides with that considered in the third row.  In this case, the
reflectance does attain the value $1$ corresponding to the band gaps,
as in the first row, since within the gaps most of the electromagnetic
energy is reflected before reaching the deep layers and thus it
doesn't sense the larger roughness.
Nevertheless, the interference oscillations of the first row
corresponding to the propagating bands are smoothed out as in the
second and third rows. On the other hand, the transmittance is very
similar to that of the third row for all frequencies as the
transmitted energy is scattered away by the rough deep
layers.
If we invert the roughness progression, the results would be similar
to those in the third row.

In the next section we will explore a real stratified system for which
the analysis of roughness that increases with depth plays a relevant
role.

\section{Application II: NAA 1D-PhC }\label{sec:AII}

We prepared an NAA 1D-PhC made of 60 repetitions of
two alternating
NAA layers of different porosities and widths on an Al
substrate by following the procedure described in
~\ref{app:AI_sample}.  We used Bruggeman's 2D model
\cite{Ortiz(2018)} to calculate the dielectric response of the porous
alumina layers in terms of their porosity and the
alumina response. We designed the system so as to have a resonance
\eqref{eq:Constructive} at the mid UV-Vis range.

\begin{figure}
  \begin{center}
  \includegraphics[width=0.9\textwidth]{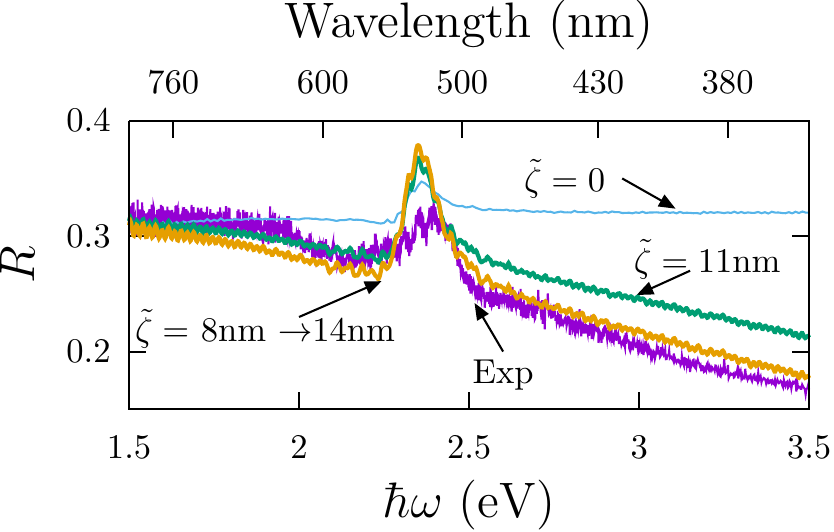}
  \caption{\label{fig:figRAP}
    Normal incidence reflectance
    $R$ as a function of photon
    energy and wavelength for a 1D-PhC made of 60 repetitions of
    two alternating film of NAA. We show
    experimental (Exp.) and several fitted theoretical
    results.
    We show results for a system with flat surfaces ($\tilde\zeta=0$), for
    a rough system with a height
    that increases linearly with depth from $\tilde\zeta=8$nm up to
    14nm, and for a rough system with constant height
    $\tilde\zeta=11$nm. We averaged $R$ over a thousand
    members of an ensemble with fluctuating layer widths characterized
    by Gaussian
    distributions with standard deviations
    $\sigma_1=\sigma_2=8$nm. In all cases we used the fitted porosities $P_1=
    29\%$ and $P_2=16\%$, and widths $d_1=90$nm and
    $d_2=87$nm.}
  \end{center}
\end{figure}

In figure \ref{fig:figRAP} we show the normal incidence reflectance
spectrum obtained experimentally together with several fitted
theoretical curves. We considered
models with flat interfaces, with constant roughness and with
linearly increasing roughness, incorporating in all cases
thickness fluctuations.
All of the fitted spectra reproduce the reflectance maximum around
2.4eV. In the absence of roughness the curve flattens out for higher
energies. Roughness yields a decrease of $R$ for frequencies above the
peak due to the extinction of the specular fields as a consequence of
scattering, but the high frequency fit is much better for the case of
progressive roughness. Such parameter optimization was performed using
the MINUIT package. The fitted parameters are the porosities
($P_1=29\%$, $P_2=16\%$), the nominal thicknesses ($d_1=90$nm,
$d_2=87$nm), the Gaussian fluctuations of the layer widths ($\sigma_1=\sigma_2=8$nm),
and the roughness height
($\tilde\zeta_1=\tilde\zeta_2=\tilde\zeta=11$nm for a constant
roughness and $\zeta=8$nm$-14$nm for linearly increasing progressive
roughness).  Since the current densities used for the 1D-PhC are those
of S2 and S3, the fitted porosities $P_2$ and $P_1$ are similar
to those reported in Table \ref{tabla1} and the fitted
thicknesses $d_i$ were in accordance to the expected ones,
given the growth conditions used. The roughness heights  are
higher than for the single slabs of Sec. \ref{sec:1slab} which
suggests that there exists an accumulation effect on the  roughness when the
unit cell is repeated several times.

\section{Conclusions}\label{sec:conclusion}

We have developed a transfer matrix formalism for stratified systems
such as 1D photonic crystals
that is capable of incorporating in a very simple way some effects of the interface
roughness. We assumed the rough surface obeys the
small angle condition. Under this
approximation \cite{Elfouhaily(2004)} the scattering matrix
\cite{Voronovich(1998)} $\mathbf S_\alpha$ of each
interface acquires a local phase that may be averaged to produce
macroscopic interfacial transfer matrices (\ref{eq:MMI}) whose
determinant is 1, and that may be incorporated into the transfer
matrix of the whole system.
The interfacial matrices
are consistent with the well
known \cite{Ogilvy(1987),Voronovich(2007)} Kirchhoff approximation
(KA), whose validity has been amply discussed
 \cite{Sanchez-Gil(1995),Pinel(2012),Tabatabaeenejad(2013),Franco(2017)}.
From the total macroscopic transfer matrix all optical properties
follow through the usual procedures for layered systems.
We extended the formalism in an appendix to the case where nearby interfaces
have mutually correlated roughness.

We employed our formalism to study a single NAA film on an Al
substrate. We verified that our assumptions held for our samples
through AFM and SEM characterizations.
We observed the characteristic oscillations
due to interference between the fields reflected by
both interfaces of the film. Roughness yields a decrease in the
reflectance for increasing frequency due to
the extinction of the specular fields through increased
scattering. This decrease is observed experimentally and is absent
from the results for flat films. Furthermore, we found a good
agreement between the parameters of the system fitted to our measurements of
the reflectance and the corresponding parameters measured directly through
SEM and AFM microscopy.

We also analyzed theoretically the photonic band structure of infinite
periodic multilayered systems. We found that roughness has an effect
similar to that of dissipation. Both blur the distinction between
propagating bands and band-gaps and produce a finite decay length
within the system, as extinction can be produced by
scattering as well as by absorption. When both absorption and roughness are
present, their effects compete and one or the other may dominate.

We studied finite slices of photonic crystals
and compared the effects of roughness to dissipation and to
fluctuations in the film widths. We found that roughness, dissipation
and fluctuations in the widths eliminate the oscillations due to the
interference between multiple reflections from the
boundaries. Furthermore, roughness and dissipation produce a similar
decrease in the reflectance maxima at the band-gaps, but width
fluctuations do not. We also studied progressive roughness that
increases towards the substrate, and which is expected in systems
such as NAA prepared by an electrochemical attack from the front
towards the back.

Finally, we applied our formalism to model the reflectance of an
NAA photonic crystal on an Al substrate and compared experimental and
theoretical results. Experiment and the theory show a maximum due to
constructive interference of fields reflected by successive periods,
but there is a decrease of the reflectance for higher frequencies
which is well reproduced by the model that incorporates roughness but
not by the model with flat surfaces.

In summary, we developed a very simple theory which allows us to
incorporate roughness into standard transfer
matrix calculations of the optical properties of stratified media. We
calculated properties of single films, of finite Bragg mirrors and of
infinite superlattices, and compared results to experiment. Our
results show that structural parameters that may be difficult to
measure directly may be obtained by analyzing optical properties and
using our formalism. Thus, we conclude that when applicable, our
formalism is useful and convenient.

\section*{Acknowledgments}
LLM is grateful to CONICET for its support through a scholarship. MLMR
is member of CONICET and acknowledges grants UBACyT
20020170200298BA and PIO 33APIO007. GPO acknowledges the support of
ANPCyT-FONCyT through grant PICT-0696-2013 and SGCyT-UNNE trough
grants PI-F008-2014 and PI-18F008. WLM acknowledges the support of
DGAPA-UNAM through grant IN111119.

\appendix
\section{Correlated rough interfaces}\label{app:correlated}

\begin{figure}
  \begin{center}
    \includegraphics[width=0.6\textwidth]{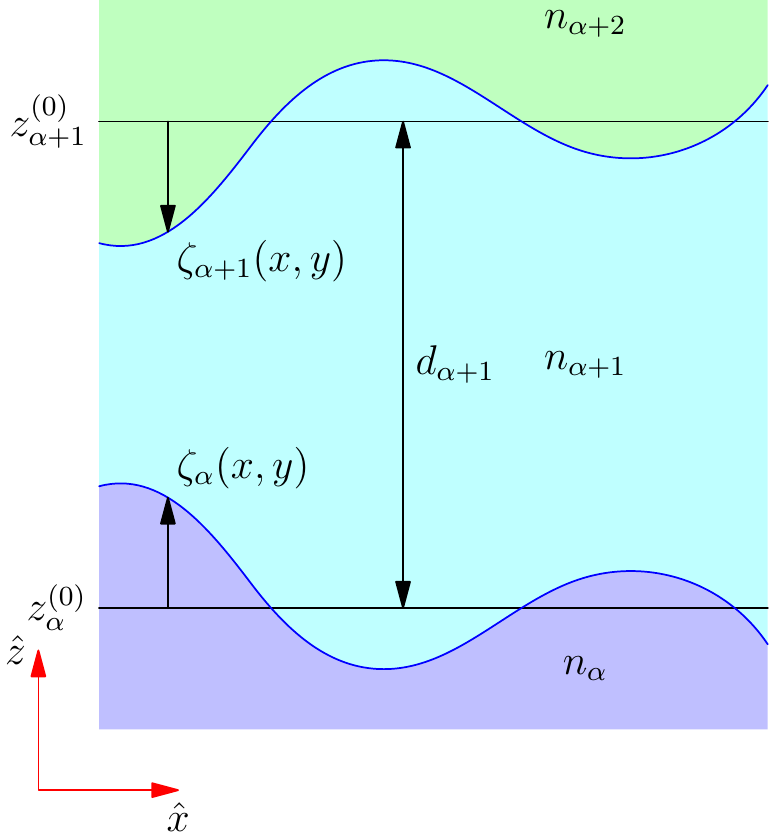}
    \caption{
      \label{fig:2perfil}
      Example of correlated rough interfaces. Layer $\alpha+1$ of
      nominal width $d_{\alpha+1}$ and refractive index  $n_{\alpha+1}$
      has perfectly
      anti-correlated interfaces at
      $z_\alpha(x,y)=z_{\alpha}^{(0)}+\zeta_{\alpha}(x,y)$
      and $z_{\alpha+1}(x,y)=z_{\alpha+1}^{(0)}+\zeta_{\alpha+1}(x,y)$, where
      $\zeta_{\alpha+1}(x,y)=-\zeta_{\alpha}(x,y)$.
    }
  \end{center}
\end{figure}

We extend the formalism of Sec. 2 to account for correlations
between two or more interfaces, as illustrated in figure~\ref{fig:2perfil}.
In case the interfaces $\alpha, \alpha+1\ldots\alpha+n$ are
correlated, we consider the corresponding contributions to the
displaced transfer matrix (\ref{eq:desplazada})
\begin{equation}\label{eq:corr}
  \mathbf M_{\alpha+n,\alpha} = \mathbf M^I_{\alpha+n} \mathbf
  M^{(0)}_{\alpha+n} \mathbf M^I_{\alpha+n-1}
  \ldots
  \mathbf M^{(0)}_{\alpha+2} \mathbf M^I_{\alpha+1} \mathbf M^{(0)}_{\alpha+1}
  \mathbf M^I_\alpha,
\end{equation}
we obtain its corresponding scattering matrix $\mathbf S_{\alpha+n,
  \alpha}$ relating the outgoing amplitudes $O^-_\alpha$ and
$O^+_{\alpha+n}$ to the incoming amplitudes $I^+_\alpha$ and
$I^-_{\alpha+n}$ by solving Eqs. analogous to Eqs. (\ref{eq:rytTEslab}) and
(\ref{eq:rytTMslab}), use them to construct the left and right matrices
$\mathbf B_{\alpha+n, \alpha}$  and $\mathbf A_{\alpha+n, \alpha}$, average them, and
use them to construct a macroscopic slab transfer matrix
\begin{equation}\label{eq:coorM}
    \mathbf M^M_{\alpha+n,\alpha} = \langle \mathbf A_{\alpha+n,\alpha}
    \rangle \langle \mathbf B_{\alpha+n,\alpha} \rangle^{-1}
\end{equation}
which may be spliced into (\ref{eq:MM}) replacing the product
$\mathbf M^{MI}_{\alpha+n} \mathbf M^{(0)}_{\alpha+n} \ldots$
$\mathbf M^{(0)}_{\alpha+1} \mathbf M^{MI}_\alpha$ to obtain the macroscopic
transfer matrix
\begin{equation}\label{eq:corrM}
  \begin{aligned}
    \mathbf M^M =& \mathbf M^{MI}_N \mathbf M^{(0)}_{N} \mathbf
    M^{MI}_{N-1}\ldots \mathbf M^{MI}_{\alpha+n+1} \mathbf
    M^{0}_{\alpha+n+1} \mathbf M^M_{\alpha+n,\alpha}
    \\
    &\mathbf  M^{0}_{\alpha} \mathbf M^{MI}_{\alpha-1}\ldots \mathbf
    M^{MI}_1 \mathbf M^{(0)}_1\mathbf M^{MI}_0,
  \end{aligned}
\end{equation}
of the rough correlated system.

\section{Experimental Section}\label{app:AI_sample}

\subsection{Materials}
Square plates of 20mm side were obtained from commercial 0.8mm thick Al
1100 \cite{Aluminio1100} sheets. Sulfuric
(H$_2$SO$_4$) and perchloric acid (HClO$_4$) (Sigma-Aldrich) were used
without further purification. Biopack absolute
ethanol and ultra-pure water were used to prepare solutions. Sintorgan
pure Acetone, Biopack Sodium hidroxide (NaOH), and Cicarelli Nitric
acid (HNO$_3$) 65\% were used for cleaning.

\subsection{NAA Sample preparation: Thin Films and 1D-PhCs}
Prior to the anodizing process, Al foils were cleaned
by  sequentially immersing and draining them in acetone for 5 min, in 10\%m/v sodium
hydroxide solution for 1 min, and nitric acid 50\%v for 1
min. Afterwords, they were thoroughly rinsed in bi-distilled water.

All samples were electroplated by setting them on a stainless steel
cathode with a contacting surface larger than the sample surface and
applying a potential of 10-12 V while immersed in a 1:4 solution of
perchloric acid and absolute ethanol. As electroplating is an
exothermic reaction, it was performed in a thermal bath with ice. The
best electroplated Al samples, with reflectance larger than 80\% and
without significant dispersion in the optical spectral range, were
obtained by fixing the voltage at 10 V during 5 min.
To prepare NAA samples we
applied a registered procedure
\cite{Calvo(2016)} with an
electrolyte composed of a solution of sulfuric acid 15\%m/v
and used a home-made
voltage and current PC controlled sources to operate the anodizing
cell~\cite{ToranzosTesis(2014)}. We have synthesized samples of
NAA films and 1D-PhC's on the Al samples. NAA thin films were prepared at
different current densities $J$ and for different times $t$ as specified in
Table~\ref{tabla1} in the manuscript. NAA 1D-PhC's were prepared by first
growing a protective thick film at 15V during 600s and then
by producing two alternating NAA layers at $J_1$=44.8mA/cm$^2$ for
$t_1$=4.5s and $J_2$=11.2mA/cm$^2$ for $t_2$=22s. This binary unit
was repeated 60 times. Following the anodizing
procedure~\cite{Calvo(2016)} to produce for each $J_i$ several NAA
films samples and by applying a linear regression of its thickness
versus anodizing times we have obtained the linear growth rates
$v_1\approx20$nm/s and $v_2\approx4$nm/s.

\subsection{Characterization Techniques}
We used an Ocean Optics USB-4000 spectrometer to measure the intensity
$I_0$ of our deuterium-tungsten-halogen light source (DT-MINI-2 Ocean
Optics), the intensity $I$ reflected at normal incidence from the NAA
films and the reference intensity $I_{\text{ref}}$ reflected from an
Al film evaporated onto a glass substrate.
We adapted a reflection probe to the spectrometer above to characterize
UV-Vis-NIR optical properties. Scanning Electron Microscopy (SEM) images were
obtained for single NAA films using a Zeiss Ultra 55 microscope. Front
views and cross-section were performed to measure porosity and
thickness respectively. Multi-mode 8 Atomic Force Microscopy (Nanoscope
V controller, Bruker, Santa Barbara) was used in contact mode to get
the NAA profiles. The cantilever spring constant was 0.32 N/m.

\section*{References}
\bibliography{referencias}

\end{document}